\documentclass[sigconf]{acmart}
\AtBeginDocument{%
  \providecommand\BibTeX{{%
    \normalfont B\kern-0.5em{\scshape i\kern-0.25em b}\kern-0.8em\TeX}}}

\setcopyright{acmcopyright}
\copyrightyear{2021}
\acmYear{2021}
\acmDOI{10.1145/3497777.3498544}

\setcopyright{othergov}\acmConference[AINTEC '21]{Asian Internet Engineering
Conference}{December 14--16, 2021}{Virtual Event, Japan}
\acmBooktitle{Asian Internet Engineering Conference (AINTEC '21), December
14--16, 2021, Virtual Event, Japan}
\acmPrice{15.00}
\acmISBN{978-1-4503-9184-9/21/12}

\usepackage{graphicx}
\usepackage{diagbox}

\begin{document}

    \title[WebRTC-based measurement tool]{WebRTC-based measurement tool for peer-to-peer applications and preliminary findings with real users}

\author{Kosuke Nakagawa}
\email{nakagawa-kosuke973@g.ecc.u-tokyo.ac.jp}
\authornotemark[1]
\affiliation{%
  \institution{The University of Tokyo}
  \streetaddress{Hongo 7-3-1}
  \city{Bunkyo}
  \state{Tokyo}
  \country{Japan}
  \postcode{113-8654}
}
\author{Manabu Tsukada}
\email{mtsukada@g.ecc.u-tokyo.ac.jp}
\orcid{0000-0001-8045-3939}
\affiliation{%
  \institution{The University of Tokyo}
  \streetaddress{Hongo 7-3-1}
  \city{Bunkyo}
  \state{Tokyo}
  \country{Japan}
  \postcode{113-8654}
}

\author{Keiichi Shima}
\email{keiichi@iijlab.net}
\orcid{0000-0003-2512-2584}
\affiliation{%
  \institution{Internet Initiative Japan}
  \city{Chiyoda}
  \state{Tokyo}
  \country{Japan}}
  
\author{Hiroshi Esaki}
\email{hiroshi@wide.ad.jp}
\affiliation{%
  \institution{The University of Tokyo}
  \streetaddress{Hongo 7-3-1}
  \city{Bunkyo}
  \state{Tokyo}
  \country{Japan}
  \postcode{113-8654}
}

\renewcommand{\shortauthors}{Nakagawa et al.}

\begin{abstract}
Direct peer-to-peer (P2P) communication is often used to minimize the end-to-end latency for real-time applications that require accurate synchronization, such as remote musical ensembles. 
However, there are few studies on the performance of P2P communication between home network environments, thus hindering the deployment of services that require synchronization.
In this study, we developed a P2P performance measurement tool using the Web Real-Time Communication (WebRTC) statistics application programming interface. Using this tool, we can easily measure P2P performance between home network environments on a web browser without downloading client applications.
We also verified the reliability of round-trip time (RTT) measurements using WebRTC and confirmed that our system could provide the necessary measurement accuracy for RTT and jitter measurements for real-time applications. In addition, we measured the performance of a full mesh topology connection with 10 users in an actual environment in Japan. Consequently, we found that only 66\% of the peer connections had a latency of 30 ms or less, which is the minimum requirement for high synchronization applications, such as musical ensembles. 
\end{abstract}

\ccsdesc[500]{Networks~Network measurement}
\ccsdesc[500]{Networks~Network performance analysis}
\ccsdesc[300]{Information systems~Web applications}
\ccsdesc[100]{Applied computing~Sound and music computing}

\keywords{Remote Collaboration, P2P, Network Measurements, WebRTC}

\maketitle

\section{Introduction}
With the new coronavirus pandemic sweeping the world, many people are now working remotely, thus increasing the opportunities for remote collaboration. For example, we currently hold remote face-to-face meetings more frequently using web conferencing applications such as Zoom.
However, for communications that require accurate synchronization, such as musical ensembles and interactive gaming, applications that use server-client communication have significant delays, which negatively affect the user experience.
To address this problem, applications that reduce latency by using peer-to-peer (P2P) to connect home networks have emerged. For example, Yamaha's Syncroom\footnote{\url{https://syncroom.yamaha.com/}} is an application that allows up to five people to play music together remotely by connecting their home networks via a P2P mesh topology connection.
However, depending on the P2P network performance, this application has not yet been widely adopted because it can cause loss of synchronization and voice disruption. It is essential to measure the actual performance of the network between homes to make the application widely used.
However, few previous studies have measured the performance of inter-home networks, and no standard measurement method has been established thus far.
This study proposes a method to measure the P2P performance between home environments using Web Real-Time Communication (WebRTC) that only requires a personal computer (PC) and a web browser.
WebRTC is an open-standard technology that provides real-time communication to browsers and mobile applications through a simple application programming interface (API).

This study aims to develop a tool to measure the performance between home networks using WebRTC without using dedicated hardware or software, and reports the results of such measurement in the case of real users.

The rest of this paper is organized as follows.
Section~\ref{sec:related} provides an overview of related works.
Section~\ref{sec:Requirements} analyses requirements for the proposed method.
Section~\ref{sec:proposed} provides the detailed description of proposed methods and implementations.
Section~\ref{sec:validation} validates our implementation. 
Section~\ref{sec:exp} describes the measurement experiment of the proposed methods.
Section~\ref{sec:discussion} discusses the results.
Section~\ref{sec:conclusion} concludes this paper.

\section{Related Works}\label{sec:related}
The overview of related works is shown in Table \ref{Comparison_of_related_work}. In this research, we aim to measure the P2P QoS of the Internet on a web browser by using WebRTC.

\begin{table*}[htb]
  \begin{center}
    \caption{Comparison of related works}
    \begin{tabular}{|c||c|c|c|}
    \hline
    \diagbox{Method}{Target}&Network Performance & WebRTC Performance & Other Application \\
    \hline\hline
    Hardware Based 
    &\cite{RIPE}\cite{Fontugne2017-lp}\cite{last-mile}\cite{sundaresan2016home}
    &\cite{barik2018can}\cite{Moulay2018-zg}\cite{tanskanen2021latency}
    &\cite{microsoft} \cite{KAIST} \\
    \hline
    Software Based
    &\cite{SINDAN}
    &\cite{Flohr2018-zf}
    &\begin{tabular}{c}
Application-specific measurement\\(Discord, Syncroom etc.)
\end{tabular}\\
    \hline
    Web Browser Based
    & \cite{mcclellan2019webrtc}, Our Research
    & \cite{Taheri2015-ho}
    &\cite{WebDINO},\cite{Garcia2017-dm}, \cite{garcia2016analysis}, Fast.com \\
    \hline
    \end{tabular}
    \label{Comparison_of_related_work}
  \end{center}
\end{table*}

There are two types of approaches for the measurement of home network environments; one relies on dedicated hardware and the other relies on dedicated software. As for those that rely on dedicated hardware, Agarwal et al. proposed a P2P latency prediction system \cite{microsoft} based on geographic information to achieve low latency matching in Halo 3, a P2P game available for Xbox 360. Youngki et al. measured the P2P communication quality for 5.6 million IP addresses playing Halo 3 on Xbox 360 and determined the round trip time (RTT) and throughput for 120 million probes in \cite{KAIST}. The authors also used the MaxMind GeoIP City Database to obtain the geographic information of the acquired IP addresses. RIPE Atlas \cite{RIPE} is a system that obtains and visualizes the latency and network outage of Internet connections at various parts of the world by deploying more than 11000 monitoring nodes worldwide. Fontugne et al. proposed a method to detect network failure points in a wide area by analyzing the measurement results by the \texttt{traceroute} command collected by RIPE Atlas in \cite{Fontugne2017-lp}. Sundaresan et al. used the measurement infrastructure of the US Federal Communications Administration to obtain TCP dump information on OpenWrt (Linux distribution) routers deployed in 2652 home networks in \cite{sundaresan2016home}. The authors created a classifier to separate upstream network congestion from home network congestion based on packet arrival and RTT variation patterns.

As for measurements that depend on dedicated software, SINDAN~\cite{SINDAN} aims to establish a method to accurately determine the network status observed from the actual environment of a user to solve claims that are often ambiguous, such as ``my network is broken,'' so that operators can find the root cause of a network issue. Home Area Latency Measurement\footnote{\url{https://iham.iijlab.net/latency/}} aimed to provide a service that estimates the latency of a home network by calculating the difference between the RTT from a home node to a measurement server and the measurement server to the exit point of the home network. Fontugne et al. revealed congestion and communication bottlenecks in the last mile by analyzing the traceroute data from RIPE Atlas in~\cite{last-mile}.

There are also web-browser-based measurement approaches. 
The iNonius project\footnote{\url{https://inonius.net/}} aims to provide a more accurate and rapid Internet measurement environment by operating an independent Internet measurement site for each organization that demands such services as telework.
WebDINO VideoMark~\cite{WebDINO} is a browser extension that measures the quality of experience (QoE) of video delivery services, such as YouTube, and is used to improve the service quality of telecommunications and video delivery service providers. 

In recent years, P2P media streaming using WebRTC has become increasingly popular, and there are many investigations on quality of service (QoS) and QoE of WebRTC applications. Moulay et al. obtained WebRTC performance from mobile nodes on MONROE, a mobile measurement platform in the European Union and evaluated the QoS and QoE in \cite{Moulay2018-zg}.
Garcia et al. 
proposed a framework that provides video quality and end-to-end latency measurement capabilities for WebRTC-based real-time application development in \cite{garcia2016analysis}.
In \cite{Garcia2017-dm}, Garcia et al.
proposed a tool to measure the QoS and QoE of WebRTC-based applications and evaluates the tool on Kurento~\cite{Garcia2017-ml}, an open-source media server compliant with the WebRTC standard.
Barik et al. 
verified that the QoS requirements of WebRTC set by applications using DiffServ Code Point work as expected in \cite{barik2018can}.
Flohr et al.
proposed a method for minimizing latency by resolving the inconsistency between the delay minimization function of real-time transport protocol and the throughput maximization function of the Stream Control Transmission Protocol (SCTP) when streaming over WebRTC in \cite{Flohr2018-zf}.
Taheri et al.
proposed a benchmarking method to measure the connection overhead and response latency of a WebRTC protocol stack implementation itself and compares the measured values against Google Chrome and Firefox in \cite{Taheri2015-ho}.
Tanskanen 
proposed a tool to explore the latency factors of WebRTC-based remote control systems in \cite{tanskanen2021latency}, implying that there is a great need for WebRTC quality measurement even in use cases of remote control.

Research has also been conducted on WebRTC-based network measurement.
In \cite{mcclellan2019webrtc}, McClellan proposed a tool to measure the network in a LAN using WebRTC.

The above-mentioned studies focus on measuring the on-premises network environment, QoE of WebRTC applications, and WebRTC-based networks in LANs. However, there has been no research on P2P RTT measurement between home environments using WebRTC.

\section{Requirements}\label{sec:Requirements}
The functional requirements of the method proposed in this study are as follows.
\begin{description}
   \item[Ease of measurement]\mbox{}\\
The potential solution must be able to measure network performance without installing dedicated hardware or software.
Measurements using RIPE Atlas and Xbox 360 depend on dedicated hardware, making it difficult to perform measurements in various environments. Although the requirements for measurements using dedicated software are less strict than those using dedicated hardware, there is still a certain amount of difficulty owing to the installation process required.  
Without the need for dedicated hardware or software, the solution can solve the problems mentioned above, and anyone can easily participate in the measurement.
   \item[Comprehensive analysis of P2P network]\mbox{}\\
It is necessary to make measurements in an environment where a mesh topology network connects multiple nodes to make measurements close to the actual usage environment. 
Real applications such as Yamaha's Syncroom enable multi-person collaboration by connecting multiple home environments through a mesh topology network. 
In addition, since a typical home environment uses a network address translation (NAT) box to connect to the Internet, it is necessary to support measurement in a NAT environment.
   \item[Intuitive visualization]\mbox{}\\
In this study, we aim to create a visualization that would allow us to intuitively grasp the trend of the RTT of P2P connections throughout the mesh topology networks. It is difficult to evaluate the statistics of a multi-node full-mesh topology connection intuitively in a raw data form. Data visualization is essential to achieve the original purpose of helping to solve P2P communication problems. 
    \item[Constrained node support]\mbox{}\\
This research aims to create a lightweight measurement tool that can be used in as many different environments as possible and run on a resource-constrained PC. By creating a tool that can run under the limitations of an old laptop computer, we can increase its accessibility and deploy our tool in more diverse home environments. Also, by creating a tool that can run on a single-board computer, such as a Raspberry Pi, it will be possible to use clients that are always connected and participating in the measurement. These nodes will increase the number of nodes participating in the measurement and keep the scale of the measurement at a certain level. 
    \item[Granularity of measurement]\mbox{}\\
    In this research, we aim to measure an RTT of less than 60 ms.
The goal of this study is to evaluate the performance of P2P connections for applications such as ensembles. Since \cite{schuett2002effects} states that the network latency that enables an ensemble to perform without problems is approximately 30 ms, we need to adopt a measurement method that can detect the fatal case of approximately 60 ms of RTT (30 ms one-way latency) to achieve the above goal.
\end{description}
By fulfilling these functional requirements, we measure RTT in more diverse P2P mesh topology connection environments on a user-centered basis without requiring dedicated hardware or software.
\section{Proposed Method}\label{sec:proposed}
\subsection{Overview}

This study proposes a method to measure the RTT of P2P connections between home environments using WebRTC.
WebRTC supports P2P communication over NAT, and we can obtain statistics such as RTT by using the \texttt{getStats()} WebRTC API method. We can use the API with a browser such as Google Chrome, which is already deployed on many PCs. 
Therefore, our WebRTC-based measurement tool fulfills the requirements described in the \textbf{ease of measurement} and \textbf{comprehensive analysis of P2P networks}. 

\subsection{Design}
\begin{figure}[tb] 
 \centering
 \includegraphics[keepaspectratio, scale=0.2]
      {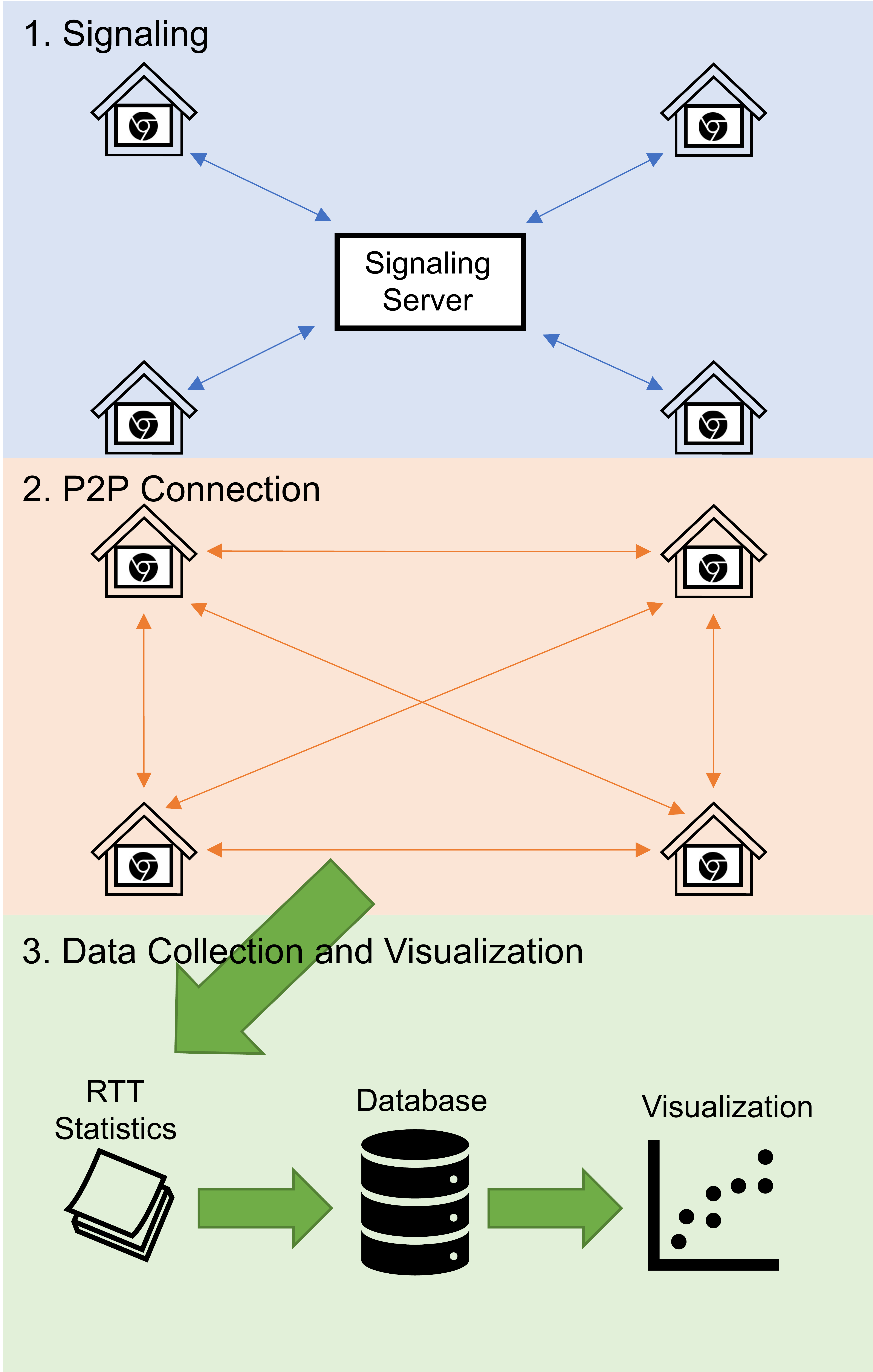}
 \caption{System Design}
 \label{design}
\end{figure}

\begin{figure*}[tb]
 \centering
 \includegraphics[keepaspectratio, scale=0.2]
      {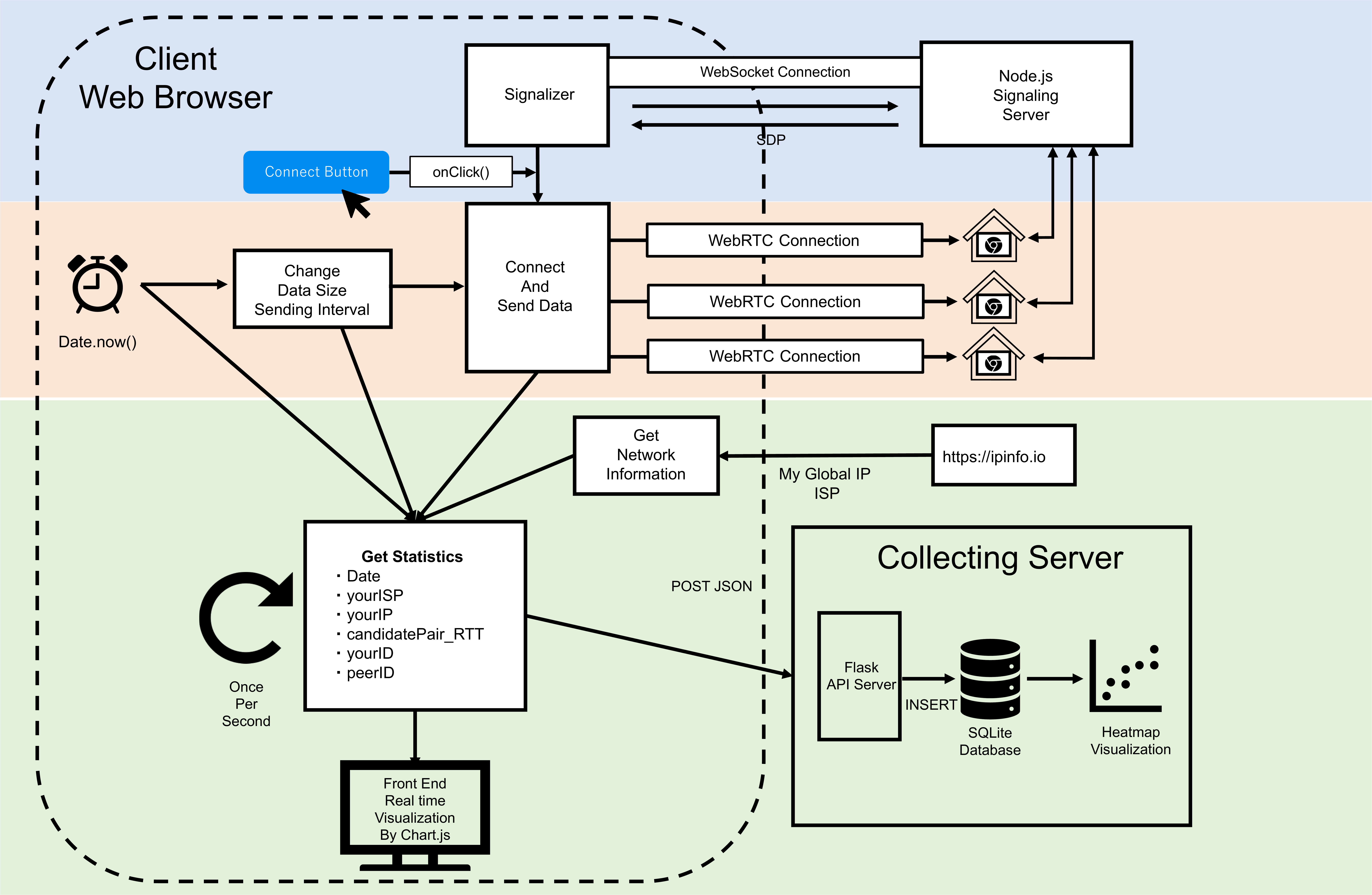}
 \caption{Implementation Overview}
 \label{implementation}
\end{figure*}

Fig. \ref{design} depicts the design architecture of this study. A user in each home network accesses the measurement website using Google Chrome, and the JavaScript measurement program runs on the browser. The operation of the measurement program can be divided into the following three stages.
\begin{itemize}
    \item Signaling
    \item P2P Connection
    \item Data Collection and Visualization
\end{itemize}
The signaling stage begins with the exchange of the Session Description Protocol (SDP) between clients. SDP is required to connect a local client and other clients connected to the measurement server over NAT boxes.
After the signaling stage is complete, the P2P connection stage is performed, which initiates the P2P connection over NAT using WebRTC and sends data at the pre-configured size and interval.
While the P2P connection stage is in progress, the data collection and visualization stage works in parallel. In the data collection and visualization stage, WebRTC regularly acquires the RTT of the P2P connection. The retrieved RTT is visualized in real-time in the front-end web browser. It is also sent to the data collection server to store the measurement data persistently in the database. The data in the database can be visualized using a heatmap after the measurement. In this way, we aim to achieve \textbf{intuitive visualization}, which is one of the requirements of this research.

\subsection{Implementation}
Fig. \ref{implementation} shows the implementation of the proposal. The measurement program for this method is a JavaScript file executed on the client's browser. When the client accesses the measurement URL, the signaling stage begins. After the signaling stage is complete, a user is required to click on the ``Connect'' button shown on the web page to establish P2P connections with other clients over NAT using WebRTC. Once P2P connections are established, the local client starts sending data to other clients with pre-configured data sizes and transmission frequencies.
While the P2P connection stage is in progress, the data collection and visualization stage is performed, which obtains the RTT, current UnixTime, and data transmission conditions every second. This statistical information is visualized in real-time on the client browser using Chart.js\footnote{\url{https://www.chartjs.org/}} and sent to the data collection server in JSON format.
The data contain the following information.
\begin{description}
    \item[Date] The result of the \texttt{Date()} call on the client web browser [ms]
    \item[yourIsp] Client ISP
    \item[yourIp] Client global IP address
    \item[candidatePair\_RTT] RTT of the P2P connection [ms]
    \item[yourID] An ID to identify the client.
    \item[peerID] An ID to identify the client.
\end{description}

The server consists of an API server using Flask\footnote{\url{https://palletsprojects.com/p/flask/}} and a database using SQLite\footnote{\url{https://www.sqlite.org/}}, and the collected data are registered in the database through the API server. The statistical information in the server database is analyzed and visualized on the server after the measurement process is completed. The statistical information that one client can obtain is limited to the P2P connection information between itself and other clients. In contrast, the server can perform data analysis in a full-mesh topology network by utilizing collected data from every client. For visualization on the server, we use techniques such as heatmaps to make the statistical information of the mesh topology networks intuitive to the user.

When sending statistics to the server, we also send the global IP address, ISP obtained by the ipinfo.io\footnote{\url{https://ipinfo.io/}} service at the beginning of the measurement in addition to the RTT of the WebRTC connection. This information is necessary for understanding the differences of RTT values between different IP addresses and ISPs.
\section{Validation of our implementation}\label{sec:validation}
In this section, we verify that the RTT of our method is negligibly different from the RTT of Ping, a well-known RTT evaluation method.
The communication protocol of WebRTC is specified by RFC 8831 \cite{RFC8831} as ``SCTP over DTLS over UDP,'' which is significantly different from conventional RTT measuring methods, such as the \texttt{ping} command. \texttt{Ping} is an RTT measurement tool that uses the Internet Control Message Protocol (ICMP)~\cite{rfc792}. In this section, we compare the RTT measurements of WebRTC with those of the \texttt{ping} command to verify the measurement result of our tool. 


We measured the RTT from an Ubuntu 18.04 PC to a Raspberry Pi 4 (2GB memory model), which were directly connected by Ethernet. Chrome and Chromium, respectively, were used as the browsers running WebRTC. Two types of measurements, WebRTC and \texttt{ping}, were performed once per second for 500 measurements.
Table~\ref{teijo} presents the results obtained when constant latencies of 0, 10, and 100 ms are enforced with the \texttt{tc} command. Regardless of the order of the latency generated by the \texttt{tc} command, the results of the WebRTC cases are 1--3 ms longer on average. 
\begin{table*}[tb]
  \begin{center}
    \caption{Comparison of WebRTC and \texttt{ping} measured round trip times (RTT)}
    \begin{tabular}{c||l|l|l|l|l|l}
    
    \hline
      \texttt{tc} command latency& \multicolumn{2}{c}{0 ms}&\multicolumn{2}{|c}{10 ms}&\multicolumn{2}{|c}{100 ms} \\
      \hline
      method & WebRTC & Ping & WebRTC & Ping & WebRTC & Ping \\
      \hline \hline
      Average RTT [ms] & 2.5 & 0.284 & 11.686 & 10.329 & 101.806 & 100.343\\
      \hline
    \end{tabular}
    \label{teijo}
  \end{center}
\end{table*}

\begin{figure}[tb]
 \centering
 \includegraphics[keepaspectratio, scale=0.55]
      {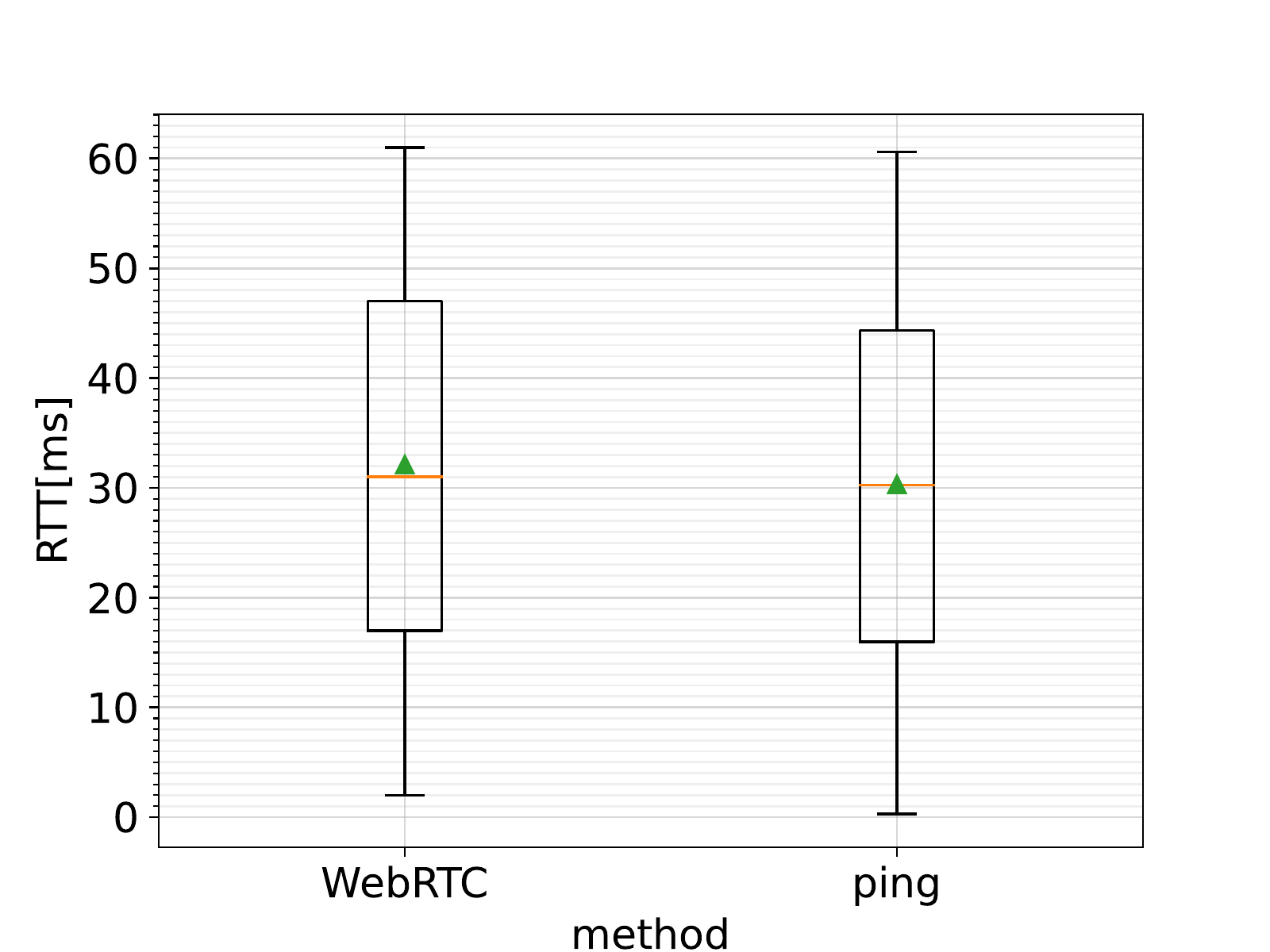}
 \caption{Measurement results with a constant latency of 30 ms and random jitter within 30 ms using the \texttt{tc} command}
 \label{jitter}
\end{figure}

Fig.~\ref{jitter} shows the boxplots of the measurement results with a constant latency of 30 ms and random jitter within 30 ms using the \texttt{tc} command. We can see 0.3--3 ms additional delay for WebRTC in all the quartiles compared to the ping results when we impose jitter. We assume that these differences are attributed to the processing overhead of WebRTC.

From these results, we can conclude there is no difference between the results of the WebRTC-based measurement and the conventional ping-based measurement except a negligible overhead seen in the WebRTC cases. We argue that the \textbf{granularity of measurement} described in Section \ref{sec:Requirements} is satisfied by the measurement using WebRTC.

\section{Experiment}\label{sec:exp}
\begin{figure*}[tb]
 \centering
 \includegraphics[width=\linewidth,clip]{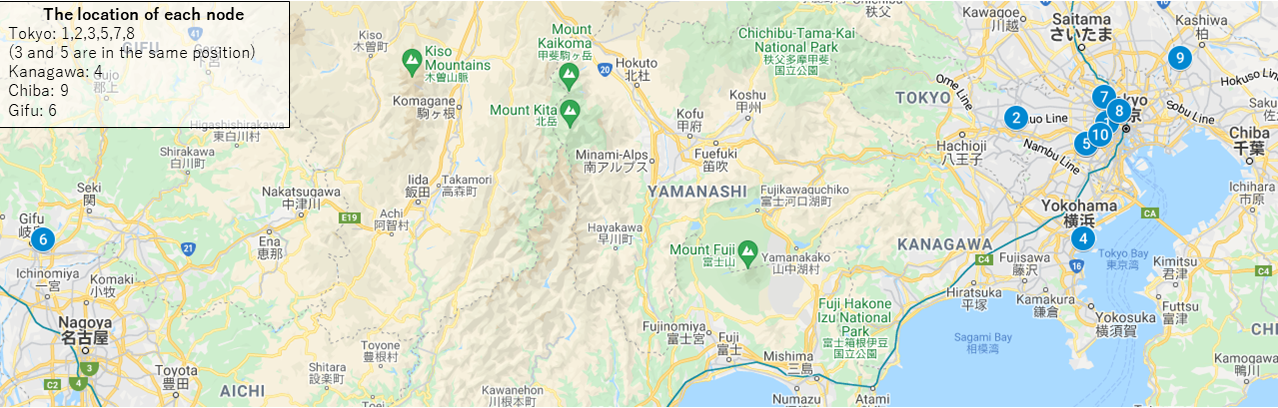}
 \caption{Geographical location of the experiment participants}
 \label{location}
\end{figure*}

\begin{figure}[tb]
 \centering
 \includegraphics[keepaspectratio, scale=0.6]
      {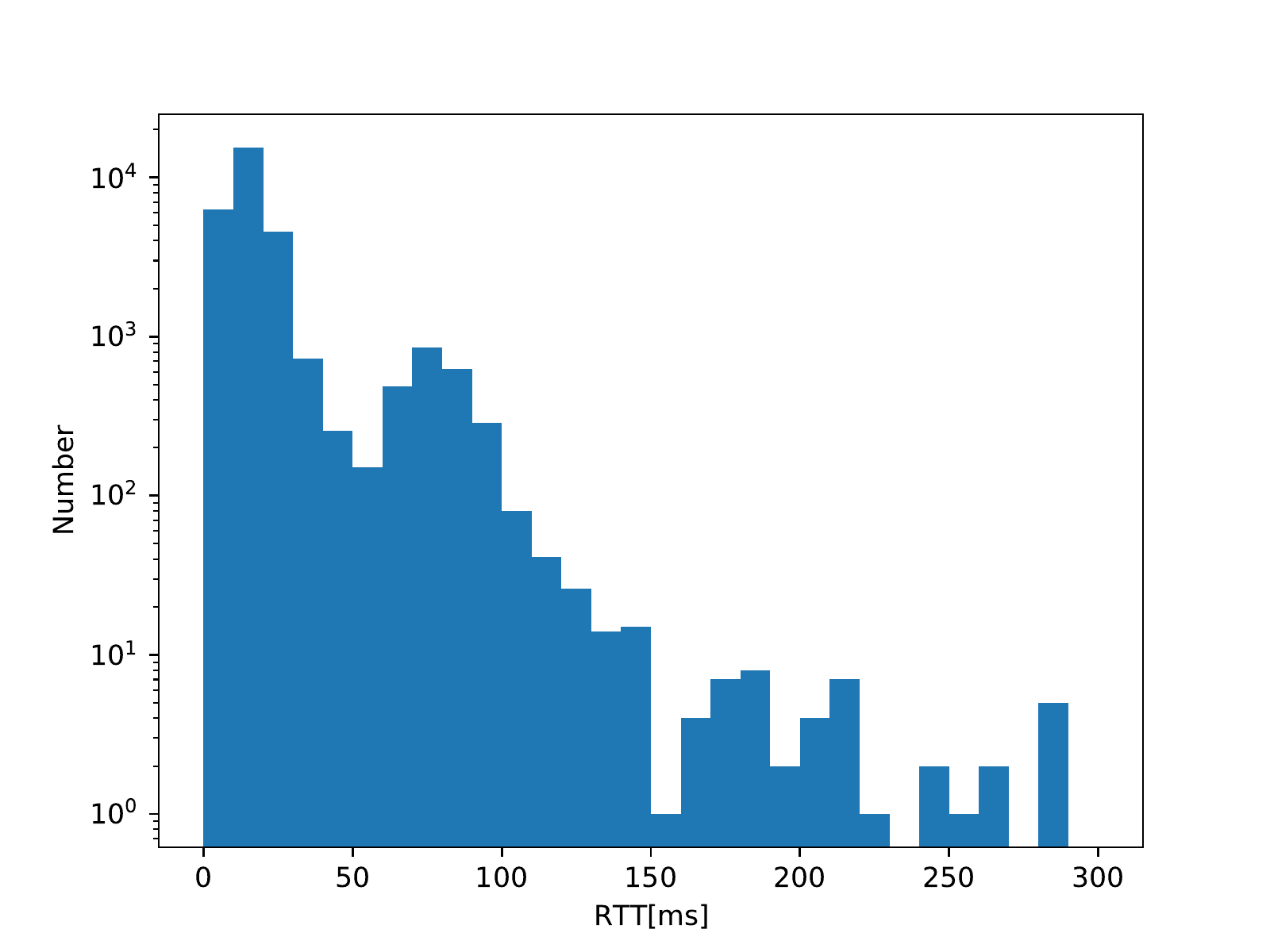}
 \caption{Distribution of the group of RTT values obtained from the entire mesh topology network}
 \label{Histogram}
\end{figure}

Using this tool, we conducted a measurement experiment to connect real home networks. We asked the experiment participants to access the measurement site using Google Chrome from their laptops in the LAN environment of their homes and organizations to perform the measurements. The number of measurement nodes used in the Kanto and Chubu regions in Japan was 10. 
Seven nodes were located in Tokyo, one in Kanagawa, one in Chiba, and one in Gifu, as shown in Fig.~\ref{location}.
During four hours from 18:00 to 22:00 (with participation and exits during this period), each client sent 100 bytes of data to the other clients and measured the RTT every second. All clients sent the measured data to the collecting server.
Fig.~\ref{Histogram} shows the distribution of the group of RTT values obtained from the entire mesh topology network. It shows that $91.7\%$ of the measurement results are between 0--60 ms, 
$7.5\%$ are between 60--100 ms, 
and only approximately $0.78\%$ measurement results are over 100 ms. Most of the measurement results meet a comfortable communication RTT of less than 60 ms, but a few high RTTs may spoil the overall experience.

\begin{figure*}[tb]
  \begin{center}
    \begin{tabular}{c}
      \begin{minipage}{0.5\hsize}
      \captionsetup{width=.8\linewidth}
          \centerline{\includegraphics[width=\linewidth,clip]{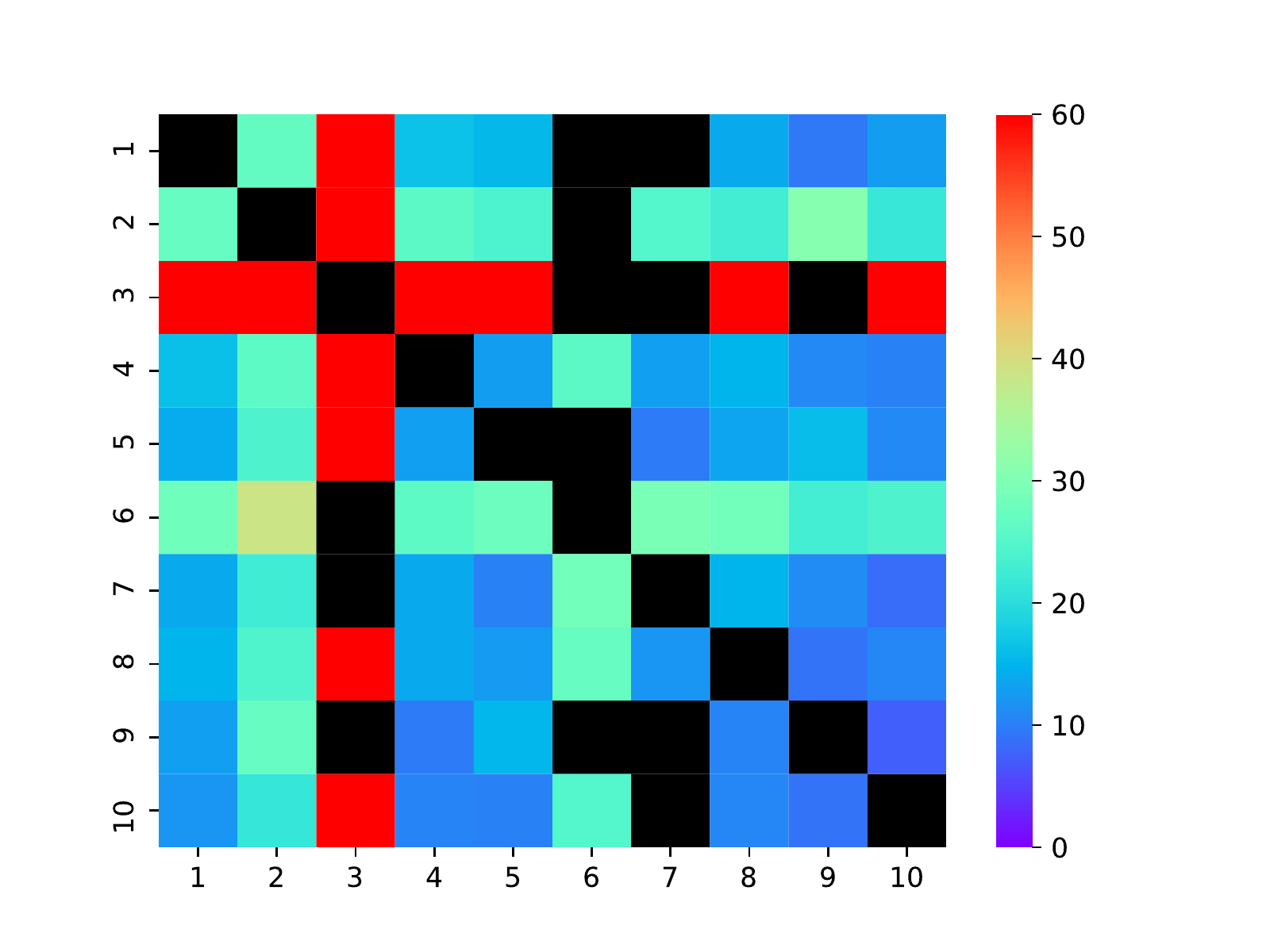}}
          \caption{Average of the group of RTT values obtained for each P2P combination [ms]}
        \label{Heatmap1}
      \end{minipage}
      \begin{minipage}{0.5\hsize}
      \captionsetup{width=.8\linewidth}
          \centerline{\includegraphics[width=\linewidth,clip]{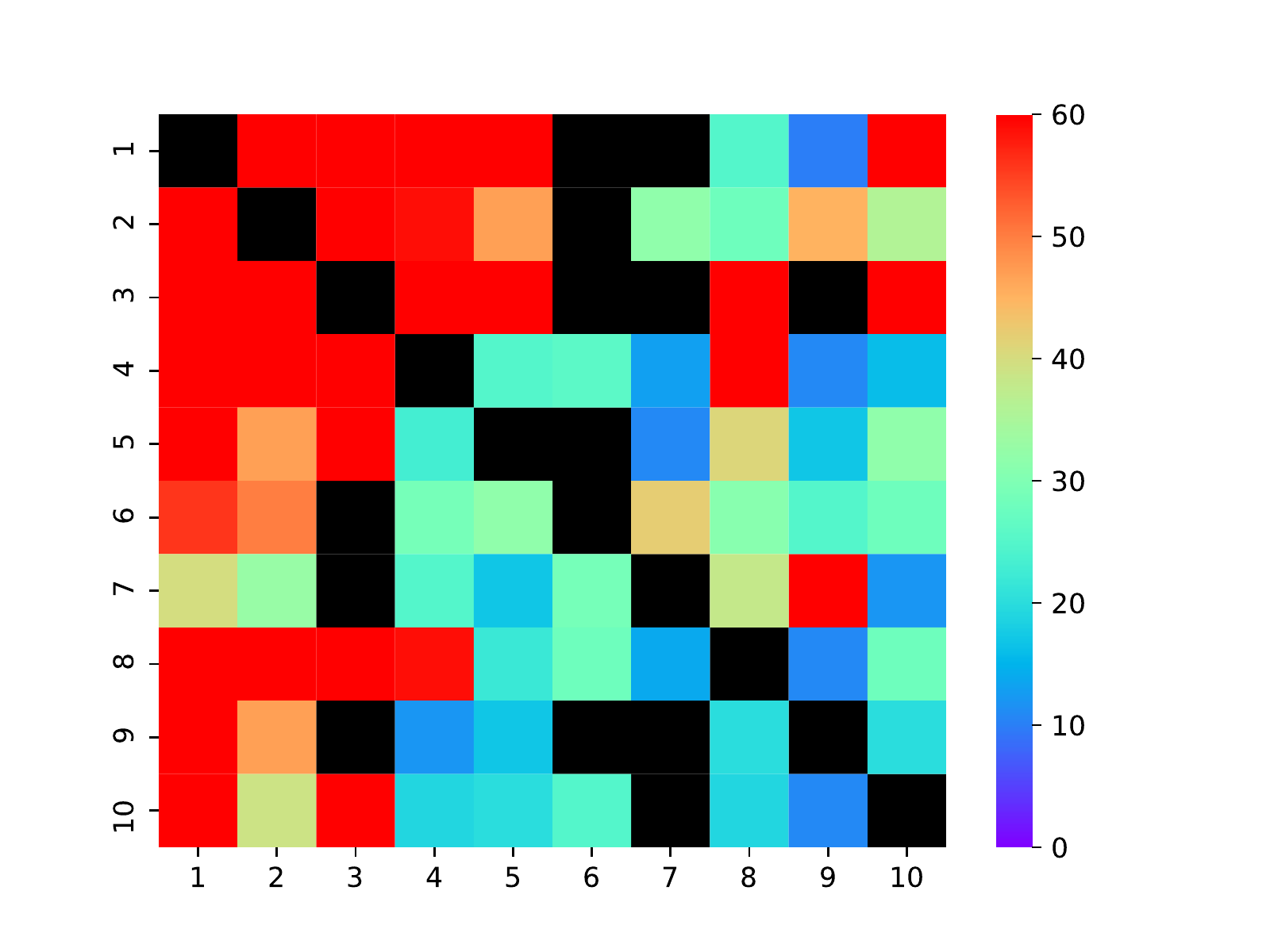}}
          \caption{99th percentile points of the group of RTT values obtained for each P2P combination [ms]}
          \label{Heatmap2}
      \end{minipage}
    \end{tabular}
  \end{center}
\end{figure*} 
The heatmaps in Fig.~\ref{Heatmap1} and Fig.~\ref{Heatmap2} illustrate the average and the 99th percentile points of the group of RTT values obtained for each P2P combination, respectively. We adopted the 99\% point to visualize the phenomenon where the RTT value increases temporarily. Even if the average latency is maintained at a low value, the QoE of the real-time communication is severely impaired when a sizeable temporary delay occurs. The label numbers in these heatmaps correspond to each measurement node and the geographical relationship in Fig.~\ref{location}. 
All the measurement results over 60 ms are indicated in red. The number of such combinations is 12 in Fig.~\ref{Heatmap1} and 26 in Fig.~\ref{Heatmap2}. 
There are three important points we can observe from the averages in Fig.~\ref{Heatmap1}. The first is that node 3 is outstandingly bad; the second is that nodes 1, 4, 5, 7, 8, 9, and 10 have low RTT, with less than 20 ms for more than half of the peers; the third is that nodes 2 and 6 are in the middle, with approximately 30 ms RTT for most of the peers, suggesting a bottleneck in the home area.

Comparing the heatmaps in Fig.~\ref{Heatmap1} and Fig.~\ref{Heatmap2} shows that nodes 1 and 2 had low RTT on average, but they had high RTT at the 99th percentile point. This result suggests that, on average, they meet the requirements for real-time applications but that sudden delays are occurring at a high frequency. If we use real-time applications in such a communication environment, the QoE is poor.
%
%


\section{Discussion}\label{sec:discussion}
In Fig.~\ref{Heatmap2}, node 1, which is located in Tokyo, has more than 60 ms of RTT with nodes 5 and 10, which are also located in Tokyo. In contrast, node 6, located in Gifu, is connected to nodes 2, 4, 5, 7, 8, and 10 in Tokyo with less than 60 ms of RTT. This result suggests that the P2P RTT between home environments is more likely to depend on home environment issues rather than geographic location. This result is understandable considering the fact that the time required to transit 270 km between Tokyo and Gifu is approximately 1 ms at the speed of light. In a country such as Japan, which is small enough not to consider the effects of the speed of light, geographic distance does not significantly affect the RTT between two home networks.

The number of peer connections with a RTT of less than 60 ms was 51 out of 77, indicating that the current Internet environment in Japan is not sufficient to allow many people to perform activities that require high synchronization, such as remote ensembles. Additionally, although the measurement results achieve less than 60 ms of RTT, this is a minimum requirement. In \cite{schuett2002effects}, a delay of 10--20 ms one way is more desirable, which is an even stricter requirement.  We expect that the environment in which we can perform satisfactory activities is even more limited.

In this study, we aimed to establish a full-mesh topology connection to emulate a realistic P2P service network. However, in reality, we could only collect RTT measurement results from 77 out of 90 peer connections in the full-mesh topology network as we could not establish a P2P connection using WebRTC in some cases. 
\section{Conclusion}\label{sec:conclusion}
In this study, we proposed and implemented a browser-based P2P measurement tool using WebRTC as a method to measure the RTT of P2P connections between home environments under various network environments. Our tool is easy to use and platform-independent. We also used this tool to obtain and visualize the RTT of P2P connections in a mesh topology with 10 nodes. As a result, we found that the number of peers capable of performing activities that require high synchronization, such as ensemble music, is very limited.

The broader implications of this research are twofold.
The first is an urgent need to develop technology to measure P2P communications' performance objectively. Remote ensemble applications using P2P connections are already widespread, but most of them operate with unknown network performance. We have presented an idea for a tool that allows end-users to measure the performance quickly.
The second is that it is significant to conduct extensive research on P2P performance in real networks.
P2P technology allows us to communicate with other terminals in the shortest path without going through a central server. This property will become increasingly important to achieve higher bandwidth and lower latency communications in areas such as edge computing, which has attracted much attention in recent years.
During the experiments of this study, we were not able to confirm the measurement in a full mesh, and some peers did not get any results. 
For some peers, only one direction was successfully measured, and we are currently investigating the cause. 
Furthermore, studying the effect of geographic distance on RTT with globally distributed peer connections is recommended.
\begin{acks}
This work was partly supported by JSPS KAKENHI (grant number: 19H04091). 
\end{acks}

\bibliographystyle{ACM-Reference-Format}
\bibliography{sample-base}

\appendix
\end{document}